\documentclass[aps,pre,floats,twocolumn,showpacs,superscriptaddress]{revtex4}

\usepackage{natbib}
\usepackage[utf8]{inputenc}

\usepackage{amsmath, amsthm, amssymb}
\usepackage{enumitem}
\usepackage{graphics,graphicx}
\usepackage{epsfig}
\usepackage{times}
\usepackage{dcolumn}
\usepackage{url}
\usepackage{color}
\usepackage{wasysym}

\usepackage[all]{xy}
\usepackage{mdwlist}

\begin{document}

\renewcommand*\thesection{\arabic{section}}
\newcommand{\beq}{\begin{equation}}
\newcommand{\eeq}{\end{equation}}
\newcommand{\sss}{\scriptscriptstyle}
\newcommand{\sz}{\scriptsize}
\newcommand{\rev}[1]{{\color{black} #1}}
\newcommand{\rerev}[1]{{\color{red} #1}}
\newcommand{\bea}{\begin{eqnarray}}
\newcommand{\eea}{\end{eqnarray}}
\newcommand{\nn}{\nonumber \\}

\title{Epidemic spreading in localized environments with recurrent mobility patterns}

\author{Clara Granell}
\affiliation{Departament de F\'isica de la Mat\`eria Condensada, Universitat de Barcelona, Mart\'i i Franqu\`es 1, E-08028 Barcelona, Spain}
\affiliation{Universitat de Barcelona Institute of Complex Systems (UBICS), Universitat de Barcelona, E-08007 Barcelona, Spain}
\affiliation{Carolina Center for Interdisciplinary Applied Mathematics, Department of Mathematics, University of North Carolina, Chapel Hill, NC 27599-3250 USA}

\author{Peter J. Mucha}
\affiliation{Carolina Center for Interdisciplinary Applied Mathematics, Department of Mathematics, University of North Carolina, Chapel Hill, NC 27599-3250 USA}

\begin{abstract}
The spreading of epidemics is very much determined by the structure of the contact network, which may be impacted by the mobility dynamics of the individuals themselves. In confined scenarios where a small, closed population spends most of its time in localized environments and has easily identifiable mobility patterns---such as workplaces, university campuses or schools---it is of critical importance to identify the factors controlling the rate of disease spread. Here we present a discrete-time, metapopulation-based model to describe the transmission of SIS-like diseases that take place in confined scenarios where the mobilities of the individuals are not random but, rather, follow clear recurrent travel patterns. This model allows analytical determination of the onset of epidemics, as well as the ability to discern which contact structures are most suited to prevent the infection to spread. It thereby determines whether common prevention mechanisms, as isolation, are worth implementing in such a scenario and their expected impact. 
\end{abstract}


\maketitle

\section{Introduction}

The spreading of infectious diseases is strongly dependent on the networked structure of interactions in the population~\cite{keeling2008modeling,pastorsatorras2015Review} and on the mobility of individuals~\cite{balcan2009Multiscale,BALCAN2010132,balcan2011Phase,brockmann2010Human,matamalas2016Assessing}. A particularly interesting scenario is that where the structure of the social contacts of the individuals is not completely well mixed nor completely structured, but offers an intermediate level of description. These kinds of models are usually referred to as metapopulation models~\cite{colizza2007Reactiondiffusion,colizza2007Invasion,colizza2008Epidemic,PhysRevE.92.062814} and were first used in the field of population ecology \cite{hanski1998Metapopulation,tilman1997spatial,bascompte1998modeling,Hanski2004xiii}. In such settings, the nodes of the network represent a population, which is occupied by individuals, and the links of the network represent the migration of individuals from one population to another. This scenario is particularly useful in the study of the spreading of epidemics, given that many real-life patterns of interactions happen in structured, localized populations connected by some degree of migration. 
The populations usually describe small, local environments (e.g., a city, a college dormitory, or an office, depending on the application) where it is plausible to assume that every individual in the population is able to contact any other individual inside the same location with some probability. The underlying network structure (i.e. the links between subpopulations) describes the mobility patterns of individuals among locations, and can be weighted and/or directed. 

The problem of modeling such scenarios relies on finding the appropriate level of abstraction to grasp the main macroscopic features of the epidemic spreading process for individuals across the particular environment. The analysis of these over-simplified model abstractions is of outmost importance to separate the effect of single parameters on the incidence of the spreading process, yet allowing an analytical approach that could be used for prediction purposes and to test prevention actions. Traditionally, models for epidemic spreading in metapopulations~\cite{colizza2007Reactiondiffusion} rely on reaction-diffusion equations to account for the epidemic and mobility dynamics, and assume that (I) individuals diffuse like random walkers through the network and (II) subpopulations with the same number of connections are treated as statistically equivalent~\cite{boguna2002epidemic}, thus smoothing over the actual contact network between individuals. While this approach has been usefully applied in many scenarios~\cite{eubank2004Modelling}, its simplified assumptions do not capture some important real-world features. For instance, analysis of human mobility data reveals that human dynamics are often dominated by recurrent patterns where individuals have memory of the location they come from \cite{rosvall2014Memory} and are highly likely to return to their original location after a short exploration of the network~\cite{balcan2009Multiscale, poletto2013human}. The typical exploration of the network mostly consists in visiting frequently a limited number of locations, predominantly performing commutes between home and work locations~\cite{belik2011Natural}. Additionally, the traditional assumption of statistical equivalence of subpopulations of the same degree, while allowing for an analytic solution of the invasion threshold, makes it impossible to quantify the outreach of an epidemic in a particular subpopulation of the network.

In this work we present a discrete-time Markov-chain model~\cite{EPL_MMCA, gomezgardenes2017Critical} for epidemic spreading in structured populations with a recurrent pattern of migrations between the locations in a bipartite network. The aim of this model is to quantify the extent of an SIS-like epidemic in the scenario where each individual spends most of their time between two locations: their \emph{residence} (e.g., home or college dormitories) and \emph{common} destinations where mixing with individuals coming from other residence subpopulations happen (e.g., work places, classes, or other common event spaces). Our goal is to discern which parameters modeling such scenarios control the phase transition of the spreading of a disease. In so doing, we can determine whether typical mechanisms of isolation---such as reducing the mobility of infected individuals---are able to contain the spreading of diseases. That is, whether or not such interventions change the critical properties of the spreading process.

The paper is organized as follows: in the next section we introduce the formulation for our model for epidemic spreading in localized environments with recurrent, bipartite travel connections. In Section 3 we show the derivation of the epidemic threshold. In Section 4 
we introduce an isolation mechanism for the infected individuals, and present the consequent formulation. Section 5 is devoted to the results of our analysis, and Section 6 offers a discussion that concludes our work.

\section{Model for epidemic spreading in metapopulations with recurrent mobility patterns} \label{sec:model}

Our metapopulation network model considers two types of locations: \emph{residences} and \emph{common} sites. Each \emph{residence} $i$ has an associated population of $n_i$ agents. Individuals associated to a given residence are assumed to interact with one other in a well-mixed fashion. A \emph{common} location, on the other hand, does not have a fixed population associated to it, thus providing a meeting site for mixing individuals from different residences. The distribution of individuals in common areas is determined by the weighted flows $W$, with elements $w_{ij}$ defining the probability of an individual associated to residence $i$ to travel to common location $j$. The flows $W$ define a bipartite network structure: no direct connections between different residences nor between different common areas are considered.

The dynamics of the model follow a discrete-time reaction-diffusion process. Every day (for each time step), individuals diffuse through the flows determined by $W$ according to the mobility probability $p$, causing $n_i p$ agents to travel to a common location and $n_i(1-p)$ individuals to remain in their residence sites, for each residence $i$. Once the individuals are in their new location, they react with the other individuals in the subpopulation (what we call the daytime infection step), meaning a susceptible individual gets infected upon contact with another infected individual with probability $\beta$. Then, agents return to their residences and another reaction is performed (nighttime infection step). After, individuals who were infected at the beginning of the time step may recover spontaneously with probability $\mu$. It is important to stress two particularities of this model. First, the daytime infection step takes place both in the common locations and in the residence sites, therefore affecting individuals who did not migrate as well as those who did.  Second, what we consider a full time step comprises two infection steps (day and night) and one recovery step.

We are interested in calculating the fraction of infected individuals assigned to any residence $i$ for each time step $t$, $\rho_i(t)$, whose time evolution is described by the following equation:
\begin{eqnarray}
\rho_i(t+1) = \rho_i(t) (1-\mu) + (1-\rho_i(t)) \Pi_i(t),  \label{eq:master}
\end{eqnarray}
The interpretation of Eq.~\eqref{eq:master} is that the fraction of infected individuals assigned to residence site $i$ at time $t+1$ is calculated as the fraction of individuals that were already infected in the previous time step and did not recover, plus those individuals who were susceptible and got infected at the end of the time step according to probability $\Pi_i(t)$, which is defined as:
\begin{eqnarray}
\begin{split}
\Pi_i(t) &=& (1-p) D_i^{\sun}(t)   +   p \sum_{j=1}^C \frac{W_{ij}}{W_i} C_j(t)  \\ &+&  p \left (   \sum_{j=1}^C \frac{W_{ij}}{W_i} (1-C_j(t))  \right )D_i^{\leftmoon}(t) \\ &+& (1-p) \left (1-D_i^{\sun}(t) \right )  D_i^{ \leftmoon}(t),  \label{eq:Pi}
\end{split}
\end{eqnarray}
\noindent where $p$ is the mobility probability and $C$ is the number of subpopulations defined as common areas. The four terms in Eq.~\eqref{eq:Pi} refer, in order, to the fraction of individuals that did not travel and got infected in their residence site in the daytime step; the fraction of people that did travel and got infected in the common site of destination; the fraction of individuals that did travel, did not get infected in the common area of destination but got infected in their residence at the nighttime step; and finally, the fraction of people that did not travel, did not get infected in their residence during the daytime step but got infected in the residence in the nighttime step. 
The expressions for the probabilities of getting infected in residence site $i$ during daytime, in residence site $i$ during nighttime and in common area $j$ are, respectively:
\begin{equation}
D_i^{\sun}(t) = 1 - (1 - \beta \rho_i(t))^{n_{i\to i}} \label{eq:Dsun} 
\end{equation}
\begin{equation}
D_i^{\leftmoon}(t) = 1 - (1 - \beta \rho_i(t))^{n_i}  \label{eq:Dmoon} 
\end{equation}
\begin{equation}
C_j(t) = 1 - \prod_{k=1}^D (1-\beta \rho_k(t))^{n_{k\to j}} \label{eq:Cj} 
\end{equation}
where $n_i$ is the size of residence $i$, $W$ is the bipartite connectivity matrix and $W_k = \sum_j^C W_{kj}$. $D$ refers to the number of residential sites. The number of individuals that remain in subpopulation $i$ is 
\begin{eqnarray}
{n_{i\to i}}= n_i (1-p), \label{eq:remain}
\end{eqnarray}
\noindent and the number of individuals moving from residence $k$ to common location $j$ is 
\begin{eqnarray}
n_{k\to j} = n_k p \frac{W_{kj}}{W_k}. \label{eq:leave}
\end{eqnarray}

\section{Calculation of the epidemic threshold} \label{sec:puntCritic}
From Eq.~\eqref{eq:master} we can calculate the solution of the system in the steady state, by assuming that $\rho_i(t+1) = \rho_i(t) = \rho_i$. Under the assumption that near the critical onset of the epidemics the fraction of infected individuals is negligible, we can substitute $\rho_i = \epsilon_i \ll 1$. Eq.~\eqref{eq:master} then reads:
\begin{equation}
\epsilon_i = \epsilon_i (1-\mu) + (1-\epsilon_i) \Pi_i.
\end{equation}
Substituting $\Pi_i$ by its expression in Eq.~\eqref{eq:Pi}, we obtain:
\begin{widetext}
\begin{equation}
\epsilon_i = \epsilon_i (1-\mu) + (1-\epsilon_i) \left[  (1-p)D_i^{\sun} + p\sum_{j=1}^C \frac{W_{ij}}{W_i} C_j + 
p \left( \sum_{j=1}^C \frac{W_{ij}}{W_i} (1-C_j)\right) D_i^{\leftmoon} + (1-p) (1-D_i^{\sun}) D_i^{\leftmoon} \right].
\end{equation}
\end{widetext}
Substituting $D_i^{\sun}$, $D_i^{\leftmoon}$ and $C_j$ by their respective expressions in Eqs.~(\ref{eq:Dsun}, \ref{eq:Dmoon}, \ref{eq:Cj}), we have:
\footnotesize
\begin{eqnarray}
\epsilon_i &=& \epsilon_i (1-\mu) + (1-\epsilon_i)  \bigl[ (1-p) \left(1-(1-\beta \epsilon_i )^{n_i(1-p)} \right)    \nn      
&+& p\sum_{j=1}^C \frac{W_{ij}}{W_i} \left(1 - \prod_{k=1}^D (1-\beta \epsilon_k)^{n_k p \frac{W_{kj}}{W_k}} \right)  \nn
&+& p \left( \sum_{j=1}^C \frac{W_{ij}}{W_i} (1-\prod_{k=1}^D (1-\beta \epsilon_k)^{n_k p \frac{W_{kj}}{W_k}})\right) \left(1 - (1 - \beta \rho_i(t))^{n_i}\right) \nn
&+& (1-p) \left(1-\left(1 - (1 - \beta \rho_i(t))^{n_i(1-p)}\right)\right) \left(1 - (1 - \beta \rho_i(t))^{n_i}\right) \Bigr]. \nonumber \\
\label{eq:gorda}
\end{eqnarray}
\normalsize

Applying the approximations $(1-\epsilon_i)^n \approx 1-n \epsilon_i$ and $\prod_{i=1}^D(1-\epsilon_i)^n \approx 1 - \sum_{i=1}^D n \epsilon_i$ and removing the $\mathcal{O}(\epsilon_{i}^2)$ terms, the previous expression reduces to:
\begin{eqnarray}
\epsilon_i &=& \epsilon_i (1-\mu) + (1-p)^2\beta n_i \epsilon_i + p \sum_{j=1}^C \frac{W_{ij}}{W_i} \beta \epsilon_i n_i \nn
&+& p^2\beta \sum_{j=1}^C \sum_{k=1}^D \frac{W_{ij}}{W_i}\frac{W_{kj}}{W_k} n_k \epsilon_k + (1-p) \beta \epsilon_i n_i.
\end{eqnarray}
We can express the previous equation in the form of an eigenvector problem, where our new expression is:
\begin{equation}
\frac{\mu}{\beta} \vec{\epsilon} = \mathbf{M} \vec{\epsilon},
\end{equation}
\noindent and thus we obtain the classical expression in epidemic spreading~\cite{wang2003epidemic}:
\begin{equation}
\beta_c = \frac{\mu}{\lambda_{\mbox{\tiny max}}(M)}, \label{eq:betacritic}
\end{equation}
\noindent where the entries of the matrix $\mathbf{M}$ are:
\begin{equation}
M_{ik} = \left( (1-p)^2n_i + n_i \right) \delta_{ik} + p^2 \sum_{j=1}^C  \frac{W_{ij}}{W_i} \frac{W_{kj}}{W_k} n_k. \label{eq:M-entries}
\end{equation}
Each entry $M_{ik}$ accounts for the average number of contacts between one individual of residence $i$ and all the individuals associated to any residence $k$ during a full day. Indeed, the first term of the r.h.s.\ of Eq.~\eqref{eq:M-entries}, accounts for the total average number of contacts among individuals of the same residence, while the second term accounts for the number of interactions that take place at the common locations.


\section{Restricting the mobility of infected individuals: the isolation factor}

Additionally, to investigate the effects of realistic isolation in our setup, we prescribe the mobility probability of infected individuals to be $p'\ll p$, thus effectively reducing their mobility through the network. The parameter that controls the relation between the two mobility rates is what we call the isolation factor $\gamma$, being $p' = \gamma p$; with $0 \le \gamma \le 1$.
This prescription changes the formulation introduced in Sec.~\ref{sec:model} as follows. First, the calculation of the number of individuals remaining in their residence (${n_{i\to i}}$)  and the number of individuals going from residence $k$ to common location $j$ ($n_{k\to j}$) need to be adjusted to take into account the two mobility probabilities. Now the probability that an individual remains in its original residential patch is $(1-\rho_i) (1-p)$ if the individual is susceptible and $\rho_i (1-p')$ if the individual is infectious. Consequently, the new expressions for equations Eqs.~(\ref{eq:remain}, \ref{eq:leave}) are:
\begin{equation}
{n_{i\to i}}= n_i [\rho_i(t)(1-p') + (1-\rho_i(t))(1-p)], 
\end{equation}
\begin{equation}
n_{k\to j} = n_k\left[\rho_k(t) p' \frac{W_{kj}}{W_k} + (1-\rho_k(t))p \frac{W_{kj}}{W_k}\right].
\end{equation}
Second, the terms $D_i^{\sun}(t)$ and $C_j(t)$ use, in the original formulation, $\rho_i$ as a proxy of the probability of infection in subpopulation $i$. This is no longer appropriate when the isolation factor is active, given that the individuals that remain in residence $i$ will no longer be an arbitrary mixing of infected and susceptible individuals. Instead, residence $i$ in the daytime step will mostly be populated by infected individuals as $p'$ grows smaller. The correct approach is to calculate the conditional probability for an individual from population $i$ to be in the infected state (I) given that the individual remains in the population during the daytime (R), which is:
\begin{eqnarray}
P(I|R)&=&\frac{P(R|I)P(I)}{P(R|I)P(I)+P(R|S)P(S)} \nn
&=&\frac{(1-p')\rho}{(1-p')\rho + (1-p)(1-\rho)}~. 
\end{eqnarray}
Using the new prescription, Eq.~\eqref{eq:Dsun} reads now:
\beq
D_i^{\sun}(t) = 1 - \left(1 - \beta \frac{(1-p')\rho(t)}{(1-p')\rho(t) + (1-p)(1-\rho(t))}  \right)^{n_{i\to i}}. \label{eq:Dsun_new} 
\eeq
Following the same rationale, we re-write the expression for Eq.~\eqref{eq:Cj}:
\beq
C_j(t) = 1 - \prod_{k=1}^D \left(1-\beta    \frac{\rho_k(t) p'}{\rho_k(t) p' + (1-\rho_k(t))p}   \right)^{n_{k\to j}}. \label{eq:Cj_new} 
\eeq
Once these changes are introduced, we can calculate the epidemic threshold of the model with isolation following the same procedure we explained in Sec.~\ref{sec:puntCritic}. After linearizing our equation and solving the eigenvector problem, we obtain the same expected expression of Eq.~\eqref{eq:betacritic}, but now the entries of matrix $\mathbf{M}$ are:
\beq
M_{ik} = \left( (1-p)(1-p')  n_i + n_i \right) \delta_{ik} + p p' \sum_{j=1}^C  \frac{W_{ij}}{W_i} \frac{W_{kj}}{W_k} n_k.
\eeq 
\noindent Note that when the isolation mechanism is not active ($\gamma = 1$), $p=p'$ and the previous expression reduces to Eq.~\eqref{eq:M-entries}. 
From the previous expression we see that the parameters that are able to shift the onset of the epidemics are the connectivity matrix $W$, the vector of sizes of the residential subpopulations $\mathbf{n}$, the mobility probability $p$ and the isolation factor $\gamma$. In the next section we explore the effects of those parameters in the final output of the epidemic process.

\section{Results}

\begin{figure}[h]
\begin{center}
  \includegraphics[width=0.95\columnwidth,clip=0]{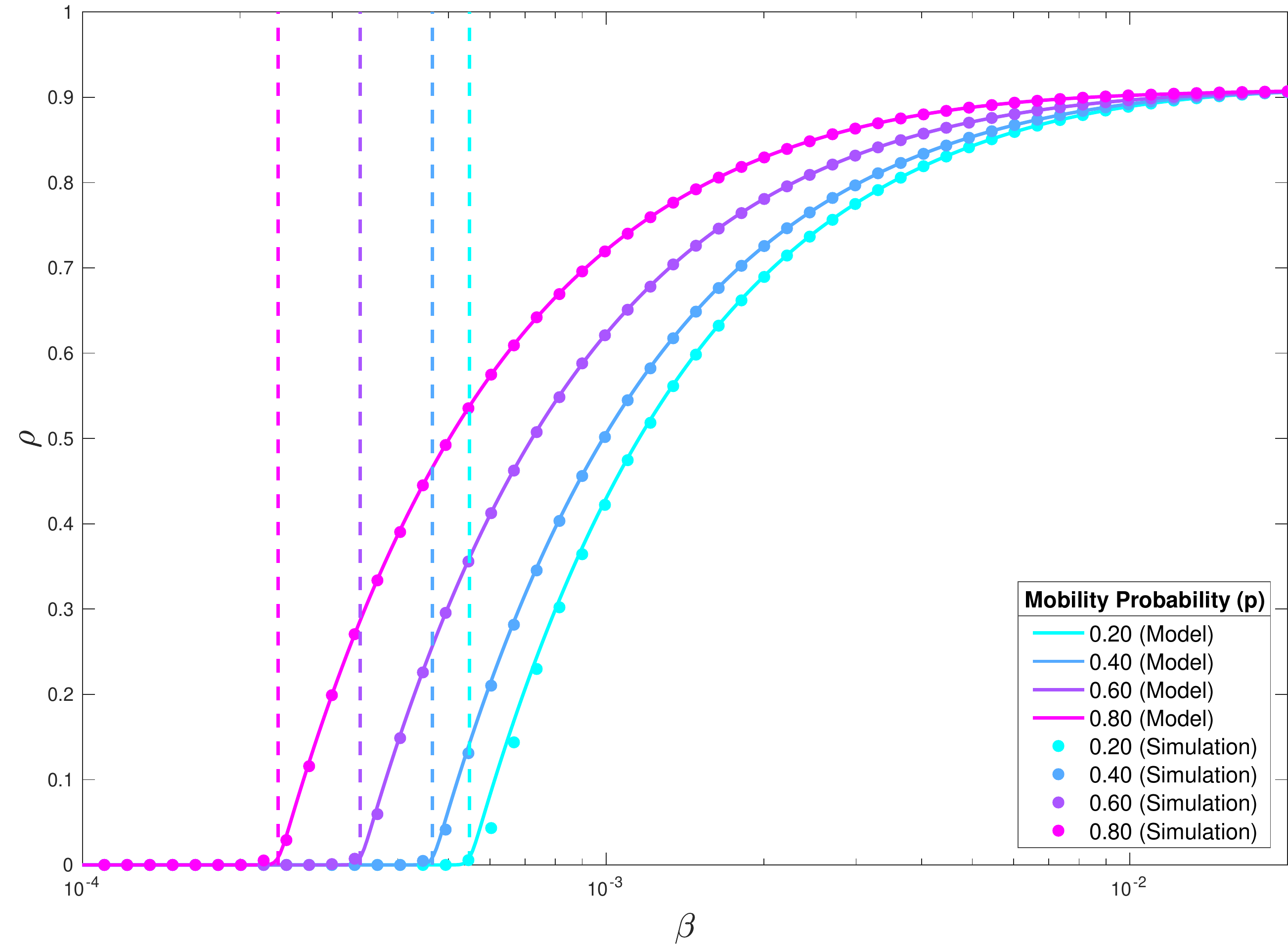}
\caption{Total fraction of infected individuals $\rho$ in the steady state as a function of the infectivity probability $\beta$, for four values of the mobility probability. Solid lines are the results of our model, while symbols are the Monte Carlo simulations. The dashed vertical lines indicate the epidemic threshold as calculated by Eq.~\eqref{eq:betacritic}. For this plot, the number of subpopulations of type residential is $D=25$ and there are $C=5$ common sites, with equal-sized residential sites of 100 individuals each. The isolation mechanism is inactive ($\gamma=1$, $p=p'$), the recovery probility $\mu=0.1$, and the connectivity matrix is an unweighted, fully connected bipartite network. }
\label{fig:montecarlo}
\end{center}
\end{figure}

To validate our model, we crosscheck the results obtained in the numerical solutions of our analytic model with extensive Monte Carlo simulations. A comparison is depicted in Fig.~\ref{fig:montecarlo}, where we plot the fraction of infected individuals in the whole system in the steady state $\rho$ as a function of the infectivity parameter $\beta$, for four values of the mobility probability $p$ and with isolation inactive. The correspondence of our analytical results with the Monte Carlo simulations is remarkable for values of the infectivity parameter even beyond the epidemic threshold. 

\begin{figure}[htbp]
\begin{center}
  \includegraphics[width=0.95\columnwidth,clip=0]{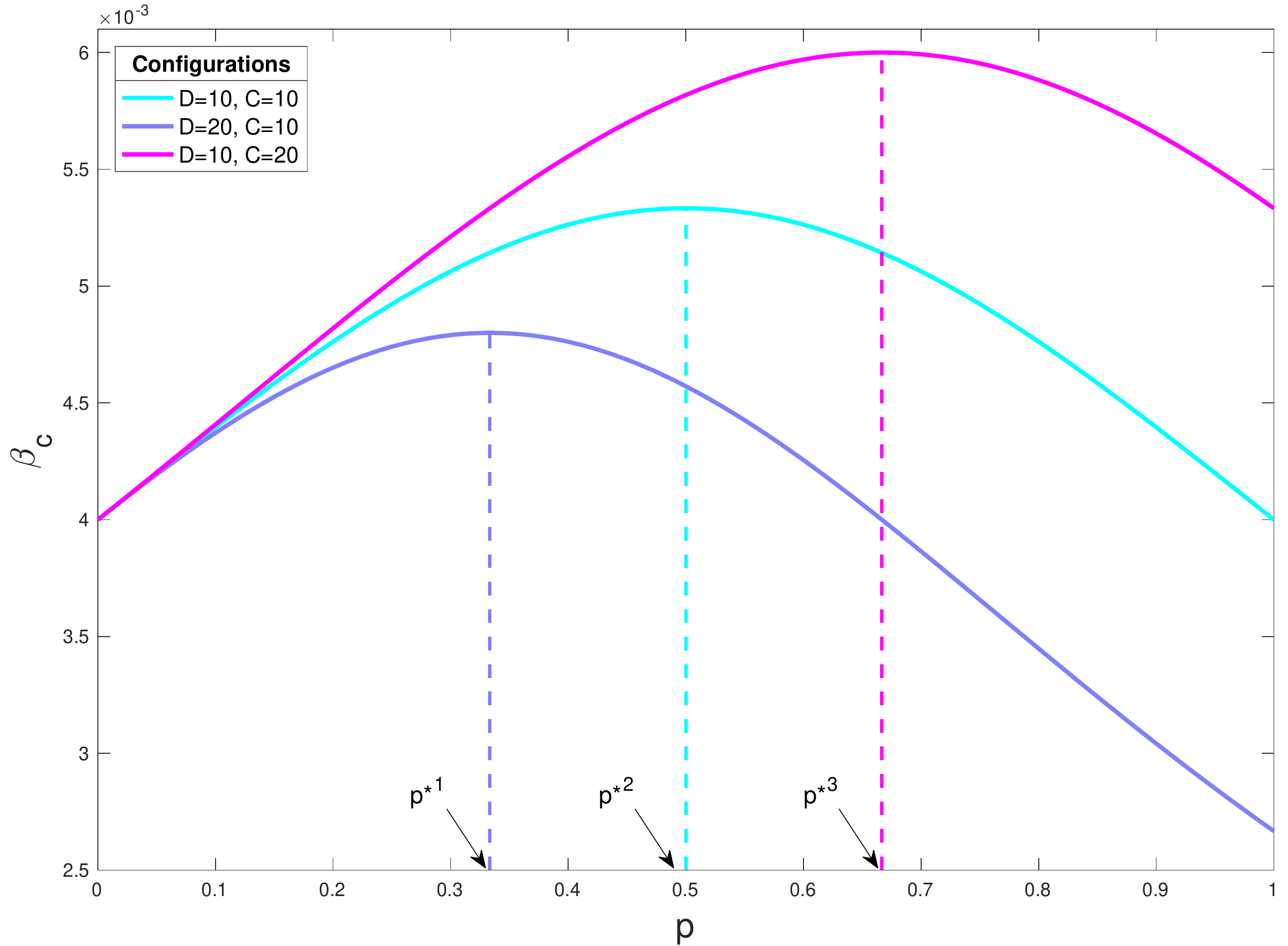}
\caption{Epidemic threshold $\beta_c$ as a function of the mobility probability $p$, for different configurations of the number of residential and common sites. We observe in all of them a non-monotonic behavior of $\beta_c$ which increases up to an optimum value of the mobility parameter ($p*$) that makes the epidemic threshold maximum (see dashed lines). Here the recovery probability $\mu=0.2$, all residential subpopulations are of the same size $n=25$, the isolation mechanism is disabled ($\gamma=1$) and the underlying topology is an unweighted fully connected bipartite network.}
\label{fig:configs}
\end{center}
\end{figure}

To analyze the effect that the mobility probability $p$ has on the epidemic threshold, we plot in Fig.~\ref{fig:configs} the curves of the critical onset of the epidemic, for different configurations on the number of residential and common sites. Here we want to highlight an interesting feature: the curve of $\beta_c$ does not have a monotonic behavior, instead there is an optimum value of the mobility probability ($p*$), which makes the epidemic threshold maximum. Indeed, we observe that $p*$ will be smaller than $0.5$ if the number of residential subpopulations exceeds the number of common sites ($D>C$); greater than $0.5$ in the opposite case ($D<C$), and exactly $0.5$ if the number of residential and common sites are equal ($D=C$), for the case of a fully connected unweighted topology and for residential sites being of the same size. This happens because $p*$ is the value of the mobility probability that causes all subpopulations in the network to be of the same (or most similar) effective size during the daytime infection step. The physics rationale of this effect can be understood as follows: the critical threshold of the epidemics is dominated by the critical threshold of the largest subpopulation, so the minimum epidemic threshold will be achieved when all populations (both residential and common sites) are of similar size. Note that the same phenomenology has been reported in \cite{gomezgardenes2017Critical} for mono-partite metapopulation networks.

\begin{figure}[btp]
\begin{center}
  \includegraphics[width=0.95\columnwidth,clip=0]{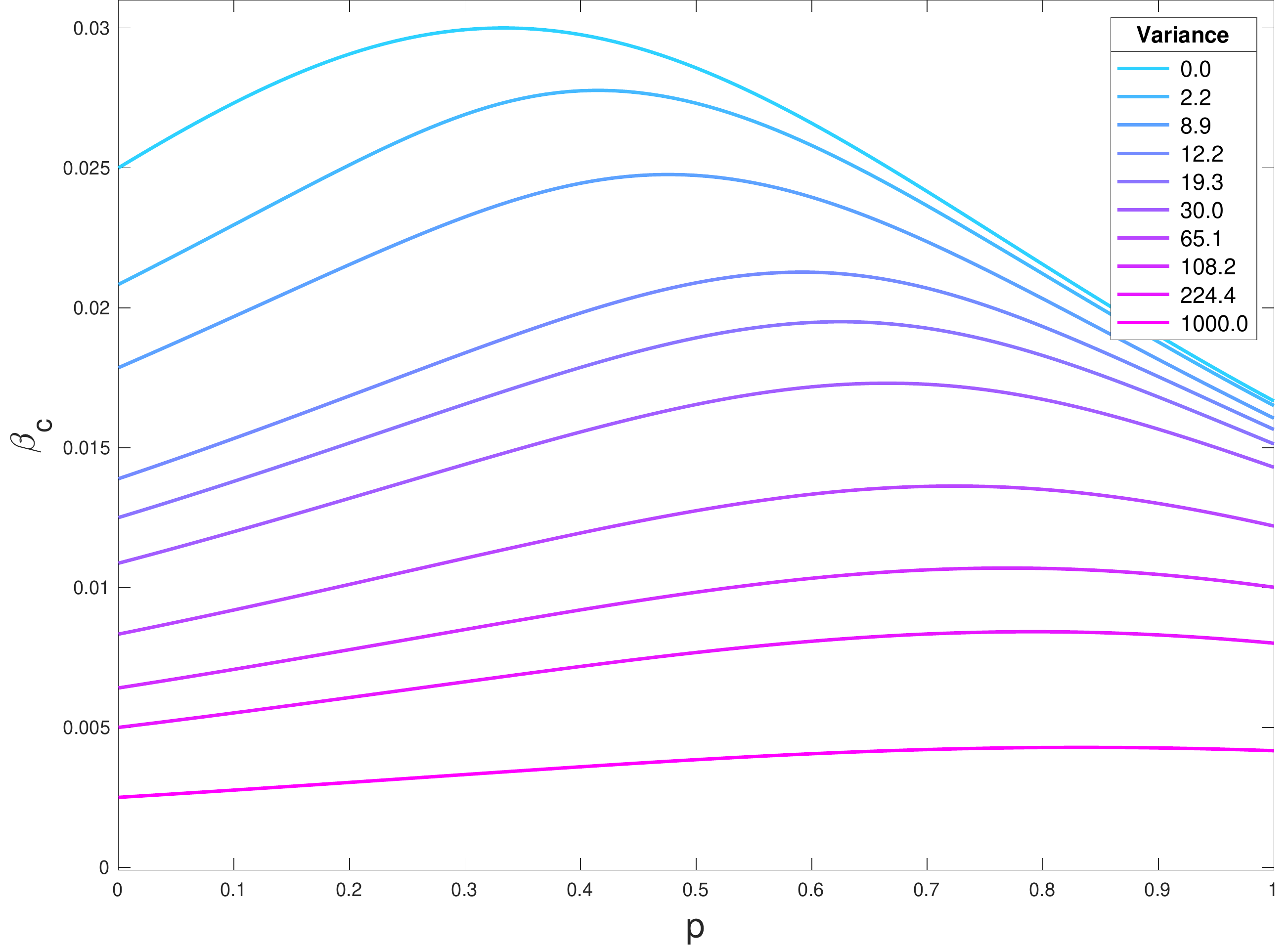}
\caption{Epidemic threshold curve as a function of mobility probability $p$, for different degrees of heterogeneity of subpopulation sizes, controlled by their variance. Here $D=10$, $C=5$ and the total number of agents is $N=100$. We vary how individuals are distributed among residential sites, ranging from the homogeneous case (all residential sites are of equal size) to the most heterogeneous setup (all individuals reside in the same node). We see that as the heterogeneity increases, the epidemic threshold gets smaller, and the effect of the optimum mobility $p*$ is diluted.}
\label{fig:vars}
\end{center}
\end{figure}

Up to now we have supposed homogeneity in the sizes of the subpopulations of type residential, meaning all entries of vector $\mathbf{n}$ are equal. Now we explore what is the effect that heterogeneity will have in the epidemic threshold. To do so, we keep the total number of individuals in the population constant, but we redistribute individuals in such a way that the variance of the size distribution increases monotonically. The results are displayed in Fig.~\ref{fig:vars}, where we observe that, as heterogeneity increases (higher variance values), values of $p*$ are shifted right and the maximum values of $\beta_c$ are less peaked. This reflects that the uneven distribution of sizes of residential subpopulations affects the critical threshold in such a way that the more heterogeneity the more easy for the epidemic to become endemic. 

\begin{figure}[tbp]
\begin{center}
  \includegraphics[width=0.95\columnwidth,clip=0]{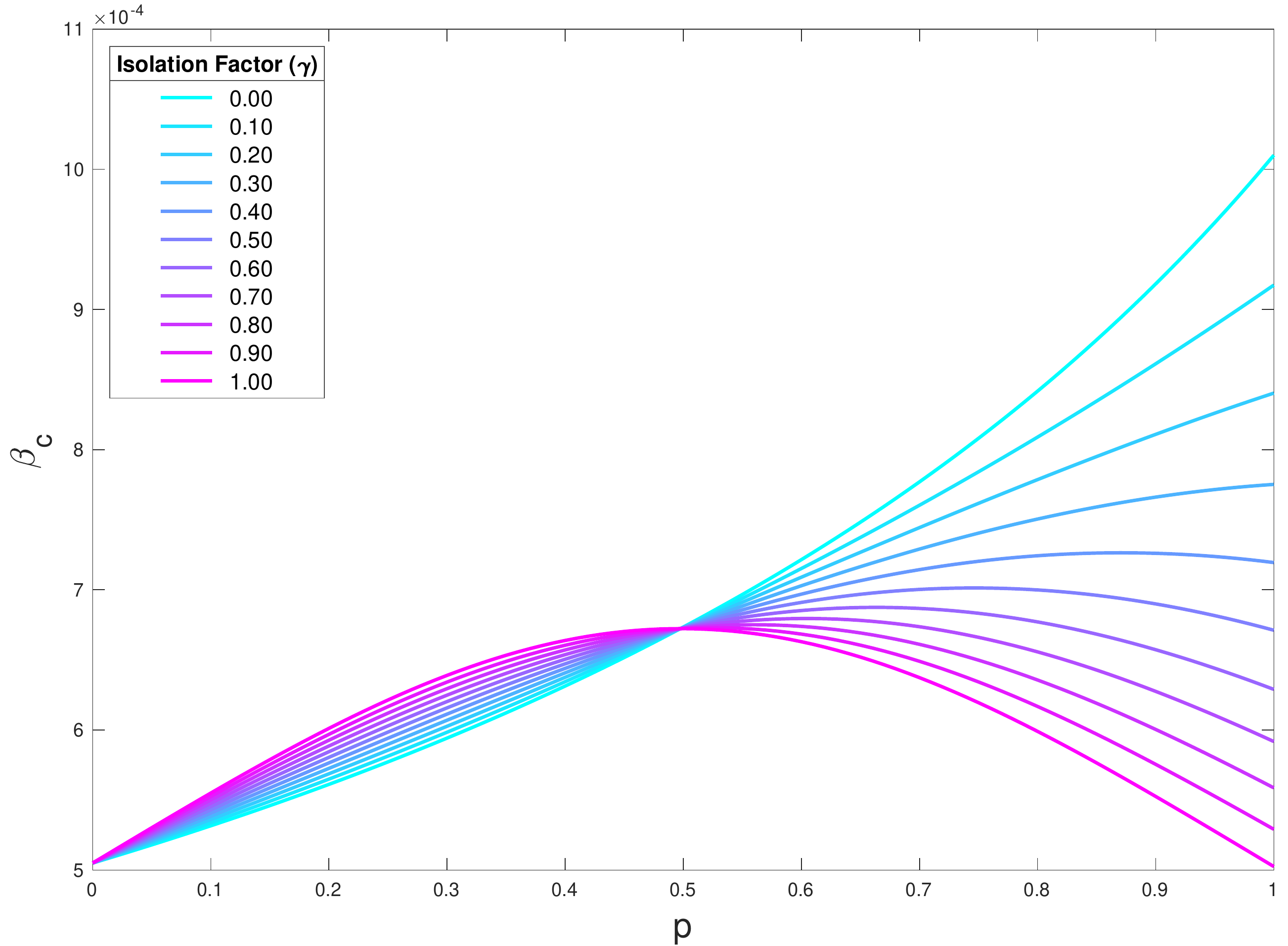}
\caption{Epidemic threshold $\beta_c$ as a function of the mobility parameter $p$, for different values of the isolation factor $\gamma$. For this plot we have used a fully connected unweighted bipartite network consisting of 10 residential sites and 10 common locations. All residential patches are of size 100 and the recovery probability is $\mu=0.1$. We observe that the curves cross at exactly the expected value $p* = 0.5$ given that there are exactly the same number of residences and common sites (see main text for a broader explanation).}
\label{fig:isolation}
\end{center}
\end{figure}

Finally, we analyze the role of the isolation factor on the critical properties of the model. In Fig.~\ref{fig:isolation} we show the epidemic threshold curve as a function of mobility probability $p$, for different settings of the isolation factor $\gamma$. We observe an interesting effect: if the isolation is inactive ($\gamma=1$), we see the increase of the epidemic threshold before $p*$ and the subsequent decrease as reported in Fig.~\ref{fig:configs}; but as we decrease $\gamma$ from $1$ to $0$ (thus gradually restricting the mobility of the infected individuals) the critical behavior of the system becomes more favorable to the epidemic extinction. As the mobility of the infected individuals is more restricted, the epidemic threshold increases with increasing mobility. In the particular example reported in Fig.~\ref{fig:isolation}, the critical behavior of the epidemic threshold is monotonically increasing for values of $\gamma\le 0.3$. We also observe that the change in the curvature of all the epidemic threshold functions coincides at exactly the expected value of $p*$, which in this case is 0.5 given that the number of residential and common sites is the same ($D$=$C$). Following the behavior observed in Fig.~\ref{fig:configs}, we also tried the configurations $D<C$ and $D>C$ and obtained that the crossing point of all curves is $p*>0.5$ and $p*<0.5$ respectively, as expected, for the case of unweighted fully connected bipartite connectivity matrices.

\section{Conclusions}

Summarizing, in this work we have proposed an analytical model to explore the spreading of epidemics in localized environments with non-random, recurrent mobility patterns. The critical properties of the epidemic process have been determined and corroborated by simulations. The results show that the main effect of the recurrent mobility is that the epidemic threshold depends on the mobility probability in a non-monotonic way, presenting an optimal value for which the epidemic is most contained. We also show that restricting the mobility of the infected individuals is an effective mechanism to to delay the critical threshold, specially for high values of the mobility. 
Importantly, the presented approach allows the appropriate modeling of epidemics on realistic scenarios that include recurrent mobility among bipartite structures, such as university campuses, home-to-work commutes or the spreading of disease in cities. The current formulation of this model is applicable to particular cases which may require locations of heterogeneous sizes, weighted connectivity and different topological structures, and allows determining whether isolation strategies are worth implementing in such specific scenarios. The presented model not only offers analytical insights to the very important problem of epidemic spreading in localized environments but could also become a powerful tool to use in data analysis and policy making. 

\section{Acknowledgements}
C.G.\ acknowledges financial support from the James S.\ McDonnell Foundation Postdoctoral Fellowship, grant \#220020457.
P.J.M.\ acknowledges financial support from the Eunice Kennedy Shriver National Institute of Child Health \& Human Development of the National Institutes of Health under Award Number R01HD075712 and from the James S.\ McDonnell Foundation under grant \#220020315. The authors acknowledge the anonymous referee for the very relevant comments provided. The content is solely the responsibility of the authors and does not necessarily represent the official views of the supporting institutions.


\end{document}